# Vibrational and electron-phonon coupling properties of $\beta$-Ga$_2$O$_3$ from first-principles calculations: Impact on the mobility and breakdown field

K. A. Mengle[1] and E. Kioupakis[1,a]

[1]*Department of Materials Science and Engineering, University of Michigan, Ann Arbor, Michigan, 48109-2136, USA*

The wide band gap semiconductor $\beta$-Ga$_2$O$_3$ shows promise for applications in high-power and high-temperature electronics. The phonons of $\beta$-Ga$_2$O$_3$ play a crucial role in determining its important material characteristics for these applications such as its thermal transport, carrier mobility, and breakdown voltage. In this work, we apply predictive calculations based on density functional theory and density functional perturbation theory to understand the vibrational properties, phonon-phonon interactions, and electron-phonon coupling of $\beta$-Ga$_2$O$_3$. We calculate the directionally dependent phonon dispersion, including the effects of LO-TO splitting and isotope substitution, and quantify the frequencies of the infrared and Raman-active modes, the sound velocities, and the heat capacity of the material. Our calculated optical-mode Grüneisen parameters reflect the anharmonicity of the monoclinic crystal structure of $\beta$-Ga$_2$O$_3$ and help explain its low thermal conductivity. We also evaluate the electron-phonon coupling matrix elements for the lowest conduction band to determine the phonon mode that limits the mobility at room temperature, which we identified as a polar-optical mode with a phonon energy of 29 meV. We further apply these matrix elements to estimate the breakdown field of $\beta$-Ga$_2$O$_3$. Our theoretical characterization of the vibrational properties of $\beta$-Ga$_2$O$_3$ highlights its viability for high-power electronic applications and provides a path for experimental development of materials for improved performance in devices.

## I. INTRODUCTION

An increasing amount of recent experimental and theoretical research has focused on the $\beta$ phase of gallium oxide ($\beta$-Ga$_2$O$_3$), with a primary focus on its applications in high-power electronic devices.[1] Because of its wider band gap and, correspondingly, its larger estimated breakdown voltage, $\beta$-Ga$_2$O$_3$ has been identified as a promising alternative for power electronics compared to other wide-gap semiconductors such as GaN and SiC.[2-7] The band gap of $\beta$-Ga$_2$O$_3$ is an especially important property to consider, with various reports placing it within a broad energy range of 4.4-5.0 eV.[2,3,8–10] Reasons for this uncertainty could include experimental growth conditions, sample type (i.e. film, bulk, etc.), sample quality, and light polarization. However, recent detailed experimental and theoretical absorption onset results point toward a fundamental band gap near 4.5 eV, i.e. at the lower limit of the reported range.[2,11,12] Although the band gap of Ga$_2$O$_3$ is lower than estimated in earlier reports, it is still much larger than other power-electronics materials, such as Si (1.1 eV), GaAs (1.4 eV), 4H-SiC (3.3 eV), or GaN (3.4 eV).[3] Since the breakdown field of a material increases strongly with increasing band-gap value, a reinvestigation of the breakdown field estimate is needed. This is achievable by studying the phonon and electron-phonon coupling properties. From the phonon dispersion, one can also obtain important information such as sound velocities and longitudinal optical (LO)-transverse optical (TO) splitting. Phonon-phonon interactions can provide information on the anharmonicity of different phonon modes, which can be used to explain finite thermal conductivity. Limited carrier mobility and breakdown field can both be explained by the electron-phonon coupling in a material. Each of these properties is important for high-power electronic applications.

---

[a] Author to whom correspondence should be addressed. Electronic mail: kioup@umich.edu.

In addition to the requirement of a wide band gap and large breakdown field, materials used in power electronics need to also exhibit large electron mobility to minimize Joule heating, as well as a high thermal conductivity to facilitate heat extraction. While the band gap and estimated breakdown field of $β$-$Ga_2O_3$ make it promising for power electronics, its electron mobility and thermal conductivity are lower than desired for such applications. An initial estimate placed the electron mobility of $β$-$Ga_2O_3$ around 300 $cm^2$ $V^{-1}$ $s^{-1}$.[3] However, both experimental and theoretical subsequent reports find much lower values. Various scattering mechanisms including acoustic deformation potential, ionized impurity, neutral impurity, and polar optical (PO) phonon scattering were analyzed alongside Hall-effect measurements by Ma *et al*. The results showed that the dominant mechanism limiting electron mobility in $β$-$Ga_2O_3$ is PO phonon scattering, which limits the room temperature mobility to < 200 $cm^2$ $V^{-1}$ $s^{-1}$ (for doping densities less than ~$10^{18}$ $cm^{-3}$).[13] Zhang *et al*. recently grew $β$-$(Al_xGa_{1-x})_2O_3$/$Ga_2O_3$ heterostructures with modulation doping, resulting in a high mobility 2D electron gas at the interface. They measured the highest experimental room temperature mobility in bulk $\beta$-Ga$_2$O$_3$ to date: 180 cm$^2$ V$^{-1}$ s$^{-1}$.[14] Two separate a*b initio* calculations using the Boltzmann transport equation (BTE) take into account scattering by PO phonons and impurities as well, resulting in room-temperature mobility values around 115 cm$^2$ V$^{-1}$ s$^{-1}$ for $n$ = 1.1×10$^{17}$ cm$^{-3}$ (Ghosh and Singisetti)[15] and 155 m$^2$ V$^{-1}$ s$^{-1}$ for $n$ = 10$^{17}$ cm$^{-3}$ (Kang *et al.*).[16] On the other hand, the measured thermal conductivity depends on the crystallographic direction due to the anisotropic monoclinic crystal structure[17], but the highest value is only 27 W m$^{-1}$ K$^{-1}$ along the [010] direction, and the lowest is 10 W m$^{-1}$ K$^{-1}$ along the [100] direction. First-principles calculations estimated the thermal conductivity along several directions (both cross-plane and in-plane) for a range of film thicknesses and found the highest value to be 21 W m$^{-1}$ K$^{-1}$ along the [010] direction.[18] To fully realize the promise of Ga$_2$O$_3$ as a high-power electronic material, the properties of phonons and their interactions with electrons and each other must be quantified to elucidate the atomistic origins of the inherent limits to its electronic and thermal transport properties.

In this work, we apply first-principles calculations based on density functional theory (DFT) and density functional perturbation theory (DFPT) to calculate the fundamental vibrational properties of $\beta$-Ga$_2$O$_3$, such as its phonon dispersion, phonon-phonon interactions, and electron-phonon coupling properties, and compare to available experimental data. Our results show that the monoclinic crystal structure strongly influences the phonon modes by causing strong anharmonicities, which are detrimental to the thermal conductivity. Moreover, our evaluated electron-phonon coupling matrix elements point to a specific low-frequency polar optical phonon mode that limits the electron mobility at room temperature. These matrix elements are also used to estimate a value of 6.8 MV/cm for the breakdown field of $\beta$-Ga$_2$O$_3$.

Our results highlight the viability of *β*-Ga$_2$O$_3$ for applications in high-power electronics and propose a path for experimental development for improved performance in devices.

This manuscript is organized as follows. In Section II we describe the computational methodology. Section III discusses the vibrational properties, including phonon frequencies, isotope effects, sound velocities, and heat capacity. Phonon-phonon interactions and their implications for thermal transport are described in Section IV, followed by a description of how the strong electron-phonon coupling in this material impacts the mobility and breakdown field in Section V. Section VI summarizes the key results of this work.

**II. COMPUTATIONAL METHODS**

Our computational methods are based on DFT and DFPT. The Quantum ESPRESSO software package was used for the calculations within the plane-wave pseudopotential formalism and the local density approximation (LDA) for the exchange-correlation potential.[19–22] Norm-conserving pseudopotentials were used for all calculations with the Ga 4*s*, 4*p*, and 3*d* and O 2*s* and 2*p* electrons included as valence states. A plane wave energy cutoff of 130 Ry and a Brillouin-zone sampling grid of 4×8×4 were determined to converge the total energy of the system to within 1 mRy/atom. The experimental lattice parameters (*a* = 12.214 Å, *b* = 3.0371 Å, and *c* = 5.7981 Å)[23] and atomic positions for *β*-Ga$_2$O$_3$ were converted to the primitive cell and used as starting points for the structural relaxation calculation. For reference, the primitive-cell lattice and the reciprocal lattice vectors are referred in this paper as $\bm{a}_1 = (-1.53, -1.46, 5.96)$, $\bm{a}_2 = (3.06, 0, 0)$, $\bm{a}_3 = (0, 5.82, 0)$, and $\bm{b}_1 = (0, 0, 0.17)$, $\bm{b}_2 = (0.33, 0, 0.08)$, and $\bm{b}_3 = (0, 0.17, 0.04)$, respectively. The structure was relaxed from the experimental lattice parameters and atomic positions until the total force on all atoms was less than $2\times10^{-5}$ Ry/$a_0$ and the total stress was less than $4\times10^{-8}$ Ry/$a_0^3$ along each direction.

Phonon frequencies $\omega(\boldsymbol{q})$ were determined with DFPT for a 4×8×4 grid of phonon wave vectors $\boldsymbol{q}$ using a stricter plane-wave cutoff energy of 170 Ry. The acoustic sum rule was applied to the phonon modes at Γ to ensure that the acoustic modes satisfy $\omega(\boldsymbol{q}) \to 0$ as $\boldsymbol{q} \to 0$. Using the dynamical matrices produced during the phonon calculations, the interatomic force constants were generated via a Fourier interpolation and applied to determine the phonon dispersion with finer sampling along each reciprocal lattice vector direction. Frequencies obtained from the interatomic force constants were compared to those obtained from an explicit phonon calculation for two q-vectors [**q** = (0, 0.25, 0) and **q** = (-1.476, 0.812, 0.329)] that are not included in the coarse grid. The difference between the interpolated and explicitly calculated frequencies agree within 1.5 cm$^{-1}$ for both q-vectors, verifying the accuracy of the Fourier interpolation. The frequency splitting between the longitudinal (LO) and transverse optical (TO) modes for infrared-active (IR) phonons along the three reciprocal lattice vector ($\boldsymbol{b_j}$) directions was calculated by including the non-analytical terms to the dynamical matrices for $\boldsymbol{q} \to 0$ along each reciprocal-vector direction. The electron-phonon coupling matrix elements between electron states $(n, \boldsymbol{k})$ and $(m, \boldsymbol{k} + \boldsymbol{q})$ are given by:

$$g_{n\boldsymbol{k},m\boldsymbol{k}+\boldsymbol{q}} = \left(\frac{\hbar}{2M\omega_i(\boldsymbol{q})}\right)^{1/2} \langle \psi_{m\boldsymbol{k}+\boldsymbol{q}} | \partial_{iq} V | \psi_{n\boldsymbol{k}} \rangle, \tag{1}$$

where $M$ is the total mass of the atoms in the unit cell, $\omega_i(\boldsymbol{q})$ is the frequency of phonon mode $i$ with wave vector $\boldsymbol{q}$ and index $i$, $\partial_{iq}V$ is the derivative of the self-consistent potential resulting from ionic displacement by phonon $i\boldsymbol{q}$, and $\psi_{n\boldsymbol{k}}$ and $\psi_{m\boldsymbol{k}+\boldsymbol{q}}$ are electronic wave functions at bands $n$ and $m$ with wave vectors $\boldsymbol{k}$ and $\boldsymbol{k}+\boldsymbol{q}$. All $g_{n\boldsymbol{k},m\boldsymbol{k}+\boldsymbol{q}}$ were determined for phonon wave vectors $\boldsymbol{q}$ along the reciprocal lattice directions by evaluating the potential derivative $\partial_{i\boldsymbol{q}}V$ for all phonon modes with DFPT and calculating the matrix element in (1) using the conduction-band electron wave functions at $\Gamma$ and at $\boldsymbol{q}$. This method for the calculation of electron-phonon coupling matrix elements yields accurate values even for the polar-optical modes for $\boldsymbol{q} \neq 0$.[24]

## III. VIBRATIONAL PROPERTIES

### A. Phonon frequencies

The calculated phonon dispersion of $\beta$-Ga$_2$O$_3$ is shown in Fig. 1. It includes thirty ($3N_{\text{atom}}$) phonon modes that span frequencies up to a maximum value of 784.92 cm$^{-1}$, which occurs at the zone edge along the $\boldsymbol{b_3}$ direction. The frequency values for the phonon modes at $\Gamma$ are listed in Table I (IR-active modes) and Table II (Raman-active modes). Directionally dependent LO-TO splitting occurs at $\Gamma$ as expected for a polar material. Table I lists both the TO and LO frequencies for each mode and compares our calculated values (obtained within LDA and Quantum ESPRESSO) to the calculated results both by Schubert et al.[8] (obtained within LDA and Quantum ESPRESSO) and Liu et al.[25] (obtained with the Perdew-Burke-Ernzerhof functional[26] and the ABINIT code[27]) and to the experimental measurements using infrared (IR) and far-infrared (FIR) generalized spectroscopic ellipsometry (GSE) by Schubert et al.[8] and infrared spectroscopic ellipsometry (IRSE) by Onuma et al.[28]

Comparing our calculated TO modes to other calculations, the majority of the frequencies agree within 5% compared to the results of Liu et al., and within 7% compared to Schubert et al.

The majority of our reported TO frequencies agree with experiment (Onuma *et al*.) better than the calculations reported by Liu *et al*., and five of our reported values are closer by more than 5% to experiment. Overall, the average error in our TO frequencies compared to the results of Onuma *et al*. is 6.3%, while the average error of the results by Liu *et al*. is slightly larger (7.9%). All but one of the TO frequencies reported by Schubert *et al*. more closely match the experimental values of Onuma *et al*. than those reported here, with their average error in TO modes being 3.6%. Experimentally determined TO modes are also reported by Schubert *et al*. Half of our calculated frequencies are in better agreement with their experimental data than their calculations, and half of their calculated values are in better agreement than ours. Our TO frequencies have an average error of 4.0%, and Schubert *et al*. have an average error of 3.4%. All three calculations represented in Table I show an IR-active phonon mode with a frequency of ~570-590 cm$^{-1}$. Schubert *et al*. also report an experimentally measured value in this range, but there is a noteworthy disagreement (~15%) between theory and experiment for this particular mode for the data reported by Onuma *et al*. who measured a value of 671 cm$^{-1}$.

We also report the LO frequencies for each IR-active mode as calculated along the three conventional crystal axes for better comparison with experiment. As expected from the crystallographic anisotropy, the LO-TO splitting depends on the direction, and some directions show larger splitting than others for a given phonon mode. The LO frequency for IR-active mode #5 agrees extraordinarily well across all reports (theoretical and experimental). A coincidental finding is that our LO frequencies agree better with the reported experimental TO frequencies than the corresponding LO frequencies for both the Schubert *et al*. and Onuma *et al*. data sets.

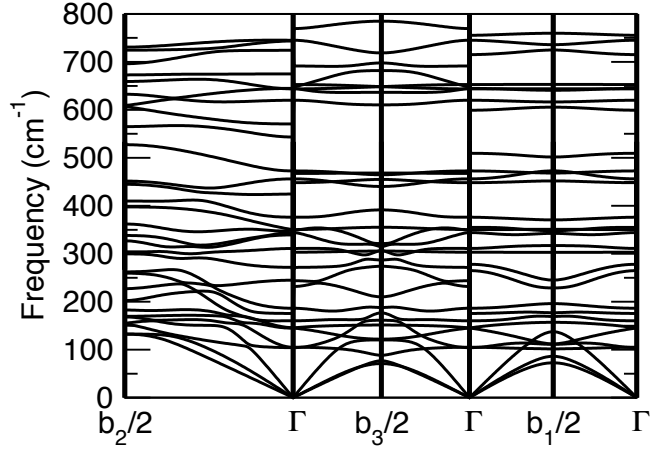

FIG. 1. The phonon dispersion of $\beta$-$Ga_2O_3$ along the directions of the reciprocal lattice vectors.

TABLE I. Calculated frequencies (in $cm^{-1}$) for IR-active phonon modes. Our values are compared to the computational results of Liu et al.[25] and Schubert et al.[8] and the experimental values measured by Schubert et al.[8] and Onuma et al.[28] Superscripts "a" and "b" correspond to the direction in which the electric field was applied in the IRSE experiments by Onuma et al.

| Mode Index | Present Work (Calc.) | | | | Liu et al.[25] (Calc.) | | Schubert et al.[8] (Calc.) | Schubert et al.[8] (Exp.) | | Onuma et al.[28] (Exp.) | |
|---|---|---|---|---|---|---|---|---|---|---|---|
| | TO | LO | | | TO | LO | TO | TO | LO | TO | LO |
| | | a | b | c | | | | | | | |
| 1 | 145.63 | 145.65 | 145.99 | 145.63 | 141.6 | 146.5 | 155.69 | 154.8 | 156.3 | – | – |
| 2 | 175.52 | 177.89 | 175.52 | 230.11 | 187.5 | 190.5 | 202.4 | 213.8 | 269 | – | – |
| 3 | 243.78 | 264.92 | 243.78 | 251.52 | 251.6 | 264.5 | 260.4 | 262.3 | 286 | 272[a] | 284[a] |
| 4 | 271.55 | 277.42 | 271.55 | 271.69 | 265.3 | 283.6 | 289.71 | 279.2 | 305 | 287[a] | 301[a] |
| 5 | 303.17 | 306.85 | 343.07 | 303.17 | 296.2 | 325.5 | 327.45 | 296.6 | 345.9 | 301[b] | 343[b] |
| 6 | 346.94 | 358.76 | 346.95 | 352.88 | 343.6 | 354.1 | 365.84 | 356.8 | 389 | 357[a] | 387[a] |
| 7 | 423.14 | 445.23 | 423.14 | 447.66 | 383.5 | 510.6 | 446.83 | 432.6 | 562 | 448[b] | 561[b] |
| 8 | 448.16 | 486.56 | 546.76 | 448.16 | 410.5 | 484.7 | 475.69 | 448.7 | 595 | 486[a] | 590[a] |
| 9 | 567.48 | 616.66 | 567.48 | 666.65 | 574.3 | 625.3 | 589.86 | 572.5 | 709 | 673[a] | 707[a] |
| 10 | 652.77 | 655.15 | 741.92 | 652.77 | 647.9 | 738.5 | 678.39 | 663.2 | 770 | 671[b] | 767[b] |
| 11 | 676.01 | 720.72 | 676.01 | 686.91 | 672.6 | 728.2 | 705.78 | 692.4 | 781.3 | 716[a] | 762[a] |
| 12 | 725.62 | 746.79 | 725.62 | 764.26 | 741.6 | 764.6 | 753.76 | 743.5 | 810 | 774[a] | 798[a] |

[a] $E||a$
[b] $E||b$

Table II lists the calculated Raman-active mode frequencies and compares them to the calculations by Liu et al.[25] and by Machon et al.[29], who both employed ultrasoft pseudopotentials within the LDA in VASP[30–32]. Experimentally obtained Raman-active phonon frequencies reported by Machon et al. are also included. The majority of our calculated frequencies agree within 3% with the other two reported sets of calculated values. Both our calculated Raman frequencies and those calculated by Liu et al. match the experimental values reported by Machon et al. nearly as well as they match each other. Only three of our calculated frequencies disagree with experiment by more than 6%. Further comparisons of experimentally and theoretically obtained phonon frequencies, are reported in Refs. 28 and 33.

TABLE II. Calculated frequencies ($cm^{-1}$) of the Raman-active phonon modes. Our results are compared to the computational work by Liu et al.[25] and to the reported computational and experimental results by Machon et al.[29]

| Mode Index | Present Work | Liu et al.[25] | Machon et al.[29] | |
|---|---|---|---|---|
| | Calculated | Calculated | Calculated | Experimental |
| 1 | 104.21 | 104.7 | 104 | 110.2 |
| 2 | 105.11 | 112.1 | 113 | 113.6 |
| 3 | 144.83 | 141.3 | 149 | 144.7 |
| 4 | 160.21 | 163.4 | 165 | 169.2 |
| 5 | 186.4 | 202.3 | 205 | 200.4 |
| 6 | 311.07 | 315.8 | 317 | 318.6 |
| 7 | 344.15 | 339.7 | 346 | 346.4 |
| 8 | 350.48 | 348.3 | 356 | – |
| 9 | 376.18 | 420.2 | 418 | 415.7 |
| 10 | 456.16 | 459.4 | 467 | – |
| 11 | 472.68 | 472.8 | 474 | 473.5 |
| 12 | 620.03 | 607.1 | 600 | – |
| 13 | 643.76 | 627.1 | 626 | 628.7 |
| 14 | 644.83 | 656.1 | 637 | 652.5 |
| 15 | 745.19 | 757.7 | 732 | 763.9 |

**B. Isotope effects on the phonon dispersion**

Previous work by Khurgin *et al.* studied the LO phonon modes of mixtures of GaN isotopes and found that isotope disorder can have a beneficial impact on materials for high-power electronic applications by increasing the density of LO phonon modes, thereby enabling more efficient cooling and an increased drift velocity.[34] Motivated by this finding and to further understand the vibrational properties of isotope-substituted $\beta$-$Ga_2O_3$, we calculated the phonon frequencies at $\Gamma$ for four different isotope substitutions ($Ga^{69}$, $Ga^{70}$, $O^{15}$, and $O^{16}$) in place of atoms with the isotopically averaged weight (i.e., 69.723 amu for Ga and 15.999 amu for O). For each comparison, we kept the mass of one atom type as its isotopic average and substituted the other with the isotopically pure value for all atoms of that type in the primitive cell.

Figure 2 shows the phonon frequencies for the TO [Fig. 2(a)], LO [Fig. 2(b)], and Raman [Fig. 2(c)] modes for each isotope variant. Reported LO values correspond to the largest LO-TO splitting out of the three reciprocal lattice vector directions. The percentile differences relative to the isotopically-averaged material are listed in Figs. 2(d)-(f). Our results show that the isotope substitution of oxygen by $O^{15}$ has the strongest effect on the phonon frequencies compared to the isotopically-averaged material. All of the high-frequency modes and approximately half of the low-frequency modes increase in frequency by about 2.5-3% in $Ga^{69.723}O^{15.000}$ compared to the average. Changing the O mass from 15.999 amu to 16.000 amu has essentially no effect on the frequencies, as is expected from such a small mass variation. Increasing the mass of Ga atoms to 70.000 amu resulted in a decrease in frequency of about 0.2% for the lower half of the phonon modes, while decreasing their mass to 69.000 amu increases the frequency of those same modes by ~0.1-0.5%.

The physical origin of the isotopic variation of the vibrational frequencies stems from the type of atoms that dominate the vibrations in each frequency range. The Ga atoms primarily contribute to many of the phonon eigenmodes at low frequencies. Since the fractional variation of the Ga mass between isotopes is small ($|\Delta m/m| \approx 1.4\%$), isotope substitution has a small effect on the frequency of those modes (up to $|\Delta\omega/\omega| = \frac{1}{2}|\Delta m/m| \approx 0.7\%$) On the other hand, the motion of the lighter O atoms dominates atomic vibrations at the high-frequency end of the spectrum and some of the low-frequency phonons. Since the fractional change of the O mass is larger ($|\Delta m/m| \approx 6.3\%$), it results in a stronger variation of the O-dominated mode frequencies ($|\Delta\omega/\omega| \approx 3.1\%$).

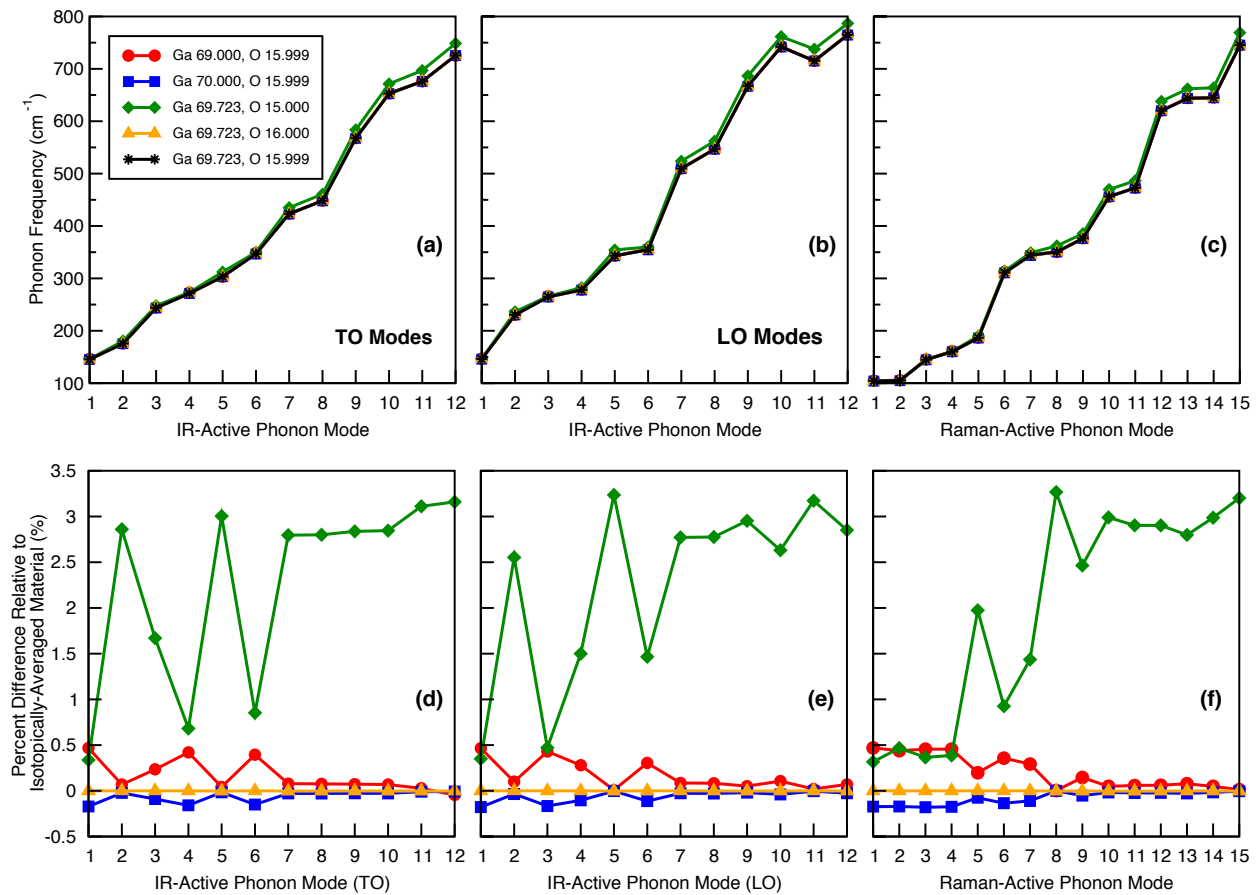

FIG. 2. Absolute [(a)-(c)] and relative [(d)-(f)] variation of the vibrational frequencies of $\beta$-Ga$_2$O$_3$ by isotopic substitution for the (a), (d) IR-active TO modes; (b), (e) IR-active LO modes with the highest frequencies out of the three reciprocal lattice vector directions; and (c), (f) Raman-active phonon modes. Substitution of oxygen by the O$^{15}$ isotope has the strongest effect on the phonon frequencies, increasing the values by up to ~3% particularly for the highest-frequency modes.

## C. Sound velocities and heat capacity

The phonon dispersion is also crucial in determining other properties of a material such as the sound velocity $v$ and heat capacity $C_V$, which are important in the evaluation of the lattice thermal conductivity. Table III shows the sound velocity of $\beta$-Ga$_2$O$_3$ along the reciprocal lattice vector directions for each of the three acoustic modes. The sound velocity takes its lowest value (2.90 km/s) for the bottom transverse acoustic (TA) mode along the $\boldsymbol{b_2}$-direction. The largest value (7.55 km/s) also occurs along the $\boldsymbol{b_2}$-direction for the longitudinal acoustic (LA) mode. In comparison, the lowest sound velocity in GaN is ~4 km/s, and the largest is ~8 km/s.[35] Since the sound velocities of Ga$_2$O$_3$ are only slightly lower than GaN, they cannot account for its much lower thermal conductivity, the origin of which will be examined later. The heat capacity at constant volume of $\beta$-Ga$_2$O$_3$ is calculated from the Helmholtz free energy $F$ according to $C_V = -T \left( \frac{\partial^2 F}{\partial T^2} \right)_V$. First, the phonon density of states (DOS) was calculated with DFPT in Quantum ESPRESSO. From the phonon DOS, the Helmholtz free energy was calculated as a function of temperature, and the second derivative with respect to temperature was taken using the finite-difference method to obtain the heat capacity. Fig. 3 shows the heat capacity at constant volume $C_V$ over a temperature range from 0-1000 K. At high temperatures, the heat capacity reaches the Dulong-Petit limit of $3N_{atom}k_B$ = 4.14×10$^{-22}$ J/K per primitive cell. Also shown in Fig. 3 is the Debye temperature ($\theta_D$) of Ga$_2$O$_3$ (738 K) which was determined from experimentally measured heat capacity data and the Debye model.[17] For comparison, GaN has a lower Debye temperature of 586 K.[36]

TABLE III. Directionally dependent sound velocities of $\beta$-Ga$_2$O$_3$ for the three acoustic phonon branches along the direction of each reciprocal lattice vector.

|  | Sound Velocities (km/s) | | |
| --- | --- | --- | --- |
|  | $\Gamma \to \boldsymbol{b_1}$ | $\Gamma \to \boldsymbol{b_2}$ | $\Gamma \to \boldsymbol{b_3}$ |
| (Bottom) TA | 3.57 | 2.90 | 2.99 |
| (Top) TA | 3.99 | 4.01 | 3.20 |
| LA | 6.18 | 7.55 | 7.46 |

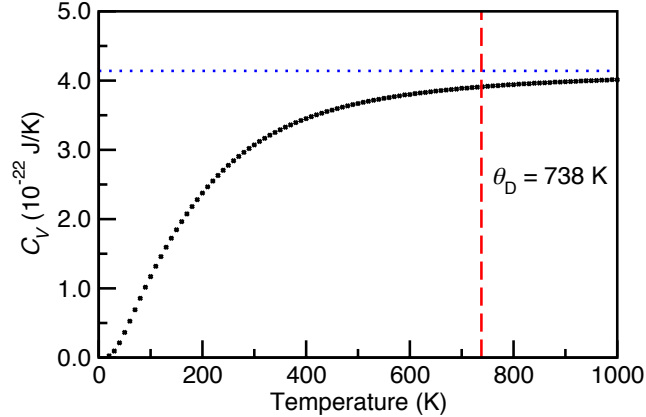

FIG. 3. The calculated heat capacity per unit cell at constant volume of $\beta$-Ga$_2$O$_3$. The heat capacity approaches the value of $3N_{atom}k_B = 4.14\times10^{-22}$ J/K per 10-atom primitive cell at high temperatures, in agreement with the Dulong-Petit law (blue, dotted horizontal line). The Debye temperature ($\theta_D = 738$ K)[17] is shown with a red, dashed vertical line.

**IV. PHONON-PHONON INTERACTIONS AND THERMAL TRANSPORT**

**A. Grüneisen parameters**

The lattice thermal conductivity of $\beta$-Ga$_2$O$_3$ at room temperature is strongly influenced by anharmonic phonon-phonon interactions, which are quantified by the Grüneisen parameters. We calculated the Grüneisen parameters for the acoustic and optical phonon modes to understand the origin of the low thermal conductivity of $\beta$-Ga$_2$O$_3$. The average Grüneisen parameter $\gamma$ for the acoustic modes is obtained by fitting the energy versus volume curve for hydrostatic deformation to the Murnaghan equation of state[37], given by:

$$E(V) = E(V_0) + \frac{KV}{K'}\left[\frac{(V_0/V)^{K'}}{K'-1} + 1\right] - \frac{KV_0}{K'-1}, \qquad (2)$$

where $E$ is the total energy of the system, $K$ is the bulk modulus, $K'$ is the pressure derivative of the bulk modulus, $V$ is the system volume, and $V_0$ is the equilibrium volume. Volume values spanning a range of ± 20% around the equilibrium were used for the fit. For hydrostatic deformation, we varied the length of the lattice vectors to obtain the desired volume while maintaining the equilibrium cell shape, and the atomic positions were relaxed to minimize the energy of the system at each volume. From fitting to the model, $K$ is determined to be 173 GPa, and $K'$ is 3.99. The best-fit curves for the Murnaghan equation of state are shown for both $\beta$-Ga$_2$O$_3$ and wurtzite GaN for comparison in Fig. 4.

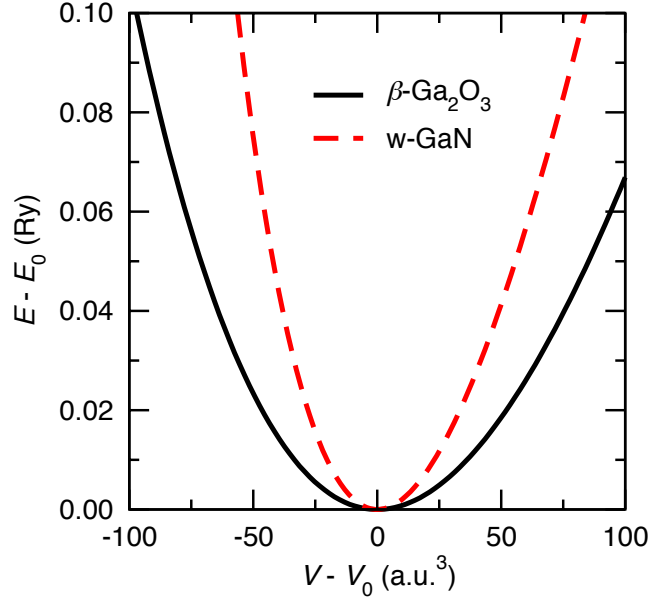

FIG. 4. Volume-dependent total system energy data was fit to the Murnaghan equation of state was for both $\beta$-Ga$_2$O$_3$ and GaN to determine the bulk modulus and pressure derivative of the bulk modulus. The curves have been rigidly shifted by $(E - E_0)$ and $(V - V_0)$. The curvature of the fitted curves show that acoustic vibrations of $\beta$-Ga$_2$O$_3$ exhibit less anharmonicity than GaN.

Once the parameter $K'$ is determined, Eq. (3) is used to calculate $\gamma$ based on the Vaschenko-Zubarev definition of the Grüneisen parameter:[38,39]

$$\gamma = \frac{K'}{2} - \frac{5}{6}. \tag{3}$$

Using the fitted $K'$ parameter from the Murnaghan equation of state, Eq. (3) was used to calculate an average acoustic Grüneisen parameter of 1.16 for $\beta$-Ga$_2$O$_3$. Another way to obtain the experimental average acoustic Grüneisen parameter is by using Eq. (4):

$$\gamma = \frac{9\alpha K V_m}{C_V}, \tag{4}$$

where $\alpha$ is the linear thermal expansion coefficient and $V_m$ is the molar volume. For comparison, applying the same methods to wurtzite GaN yields a bulk modulus of 192 GPa, pressure derivative of 4.44, and Grüneisen parameter equal to 1.39. Since the acoustic-mode $\gamma$ for GaN is larger than that for $\beta$-Ga$_2$O$_3$, yet GaN exhibits a higher thermal conductivity, we conclude that the *optical* modes of $\beta$-Ga$_2$O$_3$ must have a strong impact on limiting thermal conductivity.

For the optical modes, the mode Grüneisen parameters $\gamma_i$ were calculated according to their definition:

$$\gamma_i(\boldsymbol{q}) = -\frac{\partial \ln \omega_i(\boldsymbol{q})}{\partial \ln V}, \tag{5}$$

where $\omega_i(\boldsymbol{q})$ is the frequency of phonon mode $i$ at wave vector $\boldsymbol{q}$. The unit-cell volume was hydrostatically varied within ±1% around its equilibrium value and the atomic positions were relaxed to minimize forces on the atoms. The resulting structural parameters were subsequently used to calculate phonon frequencies.

Table IV shows the calculated Grüneisen parameters for the IR-active phonon modes. Two of these values (for the two lowest frequency IR-active modes) are negative, indicating that the corresponding modes soften (frequencies decrease) as the volume decreases. This is contrary to typical behavior where modes stiffen due to stronger bonds as the crystal volume decreases. The largest mode Grüneisen parameter is 2.32, corresponding to IR-active mode #5. Other notably large Grüneisen values for the IR modes are 1.97 (mode #9), 1.34 (mode #7), 1.28 (mode #11), and 1.25 (mode #12). Since many of the phonon modes have large Grüneisen parameters, we examined the agreement of the frequencies of these modes to experimental values and whether the frequencies are affected by strong anharmonic effects. However, closer examination of the modes with the largest (and smallest) mode Grüneisen parameters does not show a correlation between the anharmonicity of a particular mode and the magnitude of the disagreement between theory and experiment.

TABLE IV. Our calculated results for the Grüneisen parameters for the IR-active phonon modes of $\beta$-Ga$_2$O$_3$.

| IR Mode | $\gamma_i$ |
|---|---|
| 1 | -0.53 |
| 2 | -2.10 |
| 3 | 0.47 |
| 4 | 1.05 |
| 5 | 2.32 |
| 6 | 1.16 |
| 7 | 1.34 |
| 8 | 1.06 |
| 9 | 1.97 |
| 10 | 0.89 |
| 11 | 1.28 |
| 12 | 1.25 |

Grüneisen parameters for the Raman-active optical phonon modes are shown in Table V. Calculated and measured values reported by Machon *et al*. are included in Table V for comparison.[29] The majority of the Raman-mode Grüneisen parameters are larger than 1. Similar to the lowest IR-active phonon modes, the lowest two Raman modes have negative Grüneisen parameters. Raman modes #7, #8, and #12 exhibit the largest Grüneisen values (1.86, 1.71, and 1.60, respectively). We note that although the calculated Raman frequencies reported by Machon *et al*. are in better agreement with their experimentally-measured frequencies than are our calculations, our calculated Raman-mode Grüneisen parameters are in better agreement with their reported experimental $\gamma_i$ values than their calculations.

Grüneisen-parameter values around 1.0 are typically considered to be large, but several of the mode Grüneisen parameters $\gamma_i$ are significantly higher than the acoustic-mode $\gamma$ parameter for this material. This indicates that the optical modes rather than the acoustic modes are the main contributors to phonon anharmonicity in this material. $\beta$-Ga$_2$O$_3$ has several optical mode Grüneisen parameters that are larger than the single largest value for GaN, with the experimental optical mode Grüneisen parameters of GaN ranging from ~1.2-1.5.[40] Comparing the Grüneisen parameters for both materials indicates that the acoustic modes have the stronger impact on the thermal conductivity in GaN, while the optical modes are more impactful in $\beta$-Ga$_2$O$_3$.

TABLE V. Our calculated Grüneisen parameters for the Raman-active phonon modes of $\beta$-Ga$_2$O$_3$. The calculated and measured values by Machon *et al.*[29] are also listed for comparison.

| Raman Mode | Present Work | Machon *et al.*[29] Calculated | Machon *et al.*[29] Measured |
|---|---|---|---|
| 1 | -1.27 | 1.39 | – |
| 2 | -1.50 | -0.7 | – |
| 3 | 1.47 | 1.53 | 1.97 |
| 4 | 0.57 | 1.00 | 0.35 |
| 5 | 1.07 | 1.30 | 0.98 |
| 6 | 1.09 | 1.13 | 0.95 |
| 7 | 1.86 | 1.83 | 1.52 |
| 8 | 1.71 | 1.47 | – |
| 9 | 0.76 | 0.58 | 0.78 |
| 10 | 1.30 | 1.26 | – |
| 11 | 1.44 | 1.14 | 1.27 |
| 12 | 1.60 | 1.70 | – |
| 13 | 1.18 | 0.8 | 1.54 |
| 14 | 1.19 | 1.39 | 1.39 |
| 15 | 1.18 | 1.23 | 1.11 |

**B. Thermal conductivity**

The thermal conductivity is an important parameter that determines heat management in devices under high-power operation, as well as in applications such as thermoelectrics. In the simplest approximation, the lattice contribution $\kappa_L$ to the thermal conductivity of $\beta$-Ga$_2$O$_3$ is directly proportional to the values of the sound velocity and the heat capacity through $\kappa_L = 1/3\, C_V v l$, where $l$ is the phonon mean free path in the material due to anharmonic phonon scattering. The experimental values of the thermal conductivity in $\beta$-Ga$_2$O$_3$ have been measured along different crystallographic directions and vary widely due to the anisotropy of the monoclinic crystal structure; the largest value (27 W m$^{-1}$ K$^{-1}$) occurs along the [010] direction, while the [100] direction exhibits the smallest (11 W m$^{-1}$ K$^{-1}$).[41] Rigorous first-principles calculations have previously estimated the [010] direction thermal conductivity to be 21 W m$^{-1}$ K$^{-1}$.[18] These thermal-conductivity values are much lower than those of other power electronic materials such as GaN (~130 W m$^{-1}$ K$^{-1}$) and SiC (~500 W m$^{-1}$ K$^{-1}$).[42] We attribute the low thermal conductivity of $\beta$-Ga$_2$O$_3$ to the large values for $\gamma_i$ (e.g., for IR-active modes #5, #9, #7, #11, and #12 and Raman-active modes #7, #8, and #12) and, in particular, to IR-active mode #5 ($\omega = 303.17$ cm$^{-1}$) with the largest mode Grüneisen parameter of 2.32.

**V. ELECTRON-PHONON COUPLING, MOBILITY, AND DIELECTRIC BREAKDOWN**

**A. Electron-phonon coupling matrix elements**

We studied electron-phonon interactions in β-Ga$_2$O$_3$ to gain a fundamental understanding of how they impact the electron transport properties relevant for power electronics. We evaluated the electron-phonon coupling matrix elements between states at Γ and at several points along each reciprocal lattice direction in increments of $0.1|b_j|$. Fig. 5 shows these electron-phonon coupling matrix elements for the bottom conduction band as a function of the phonon wave vector. Fig. 5 also displays the magnitude of the phonon-absorption ($g^2(q)n_q$) and phonon-emission ($g^2(q)(n_q + 1)$) contributions of each phonon mode to the mobility, where $n_q$ are the phonon occupation numbers given by the Bose-Einstein distribution:

$$n_q = \left( e^{\frac{\hbar\omega(q)}{k_B T}} - 1 \right)^{-1} \qquad (6)$$

evaluated at 300 K ($k_B T = 25.85$ meV). For clarity, the plots in Fig. 5 only include the modes that have the top three largest values for either the matrix element squared, phonon-absorption, or phonon-emission terms along each direction. Note that the phonon frequencies given in Fig. 5 correspond to the frequencies of the modes at the calculated wave vectors $q$ closest to, rather than at, Γ since these are the specific wave vectors used in Eqs. (6) and (7).

As in polar materials in general, the phonon modes that dominate electron-phonon coupling in β-Ga$_2$O$_3$ are the polar LO phonon modes[43] and show a diverging behavior as $q \to 0$. The interaction of electrons with the polar LO phonons is described by the Fröhlich expression of the general form:

$$g_{i,b_j}^{2}(\mathbf{q}) = C_{i,b_j}/q^2 \qquad (7)$$

for each phonon mode $i$ and wave vector $\mathbf{q}$ along each reciprocal lattice vector direction $\mathbf{b_j}$. We determined the numerators $C$ from the calculated matrix elements for the wave vectors closest to $\Gamma$ along each direction. Table VI contains the calculated values of $C$ for the top several phonon modes with the largest phonon-absorption and emission values. The $C$ values are used to estimate the breakdown field of $\beta$-Ga$_2$O$_3$ in Section V.C.

The majority of the dominant phonon modes with respect to electron-phonon interactions have high phonon frequencies. Among them the mode with the lowest frequency has a value of 235 cm$^{-1}$. Notably, the energy of this phonon mode (29 meV) is very near that of $k_BT$ at room temperature (26 meV), which indicates that this mode has a high occupation number at room temperature. For all three directions, the highest frequency modes have the largest Fröhlich model $C$ coefficients, and the largest $C$ values occur along reciprocal lattice directions $\mathbf{b_3}$ and $\mathbf{b_1}$ (1091 meV$^2$/a$_0^2$ and 1074 meV$^2$/a$_0^2$, respectively). The coefficient values tend to decrease with decreasing phonon frequency.

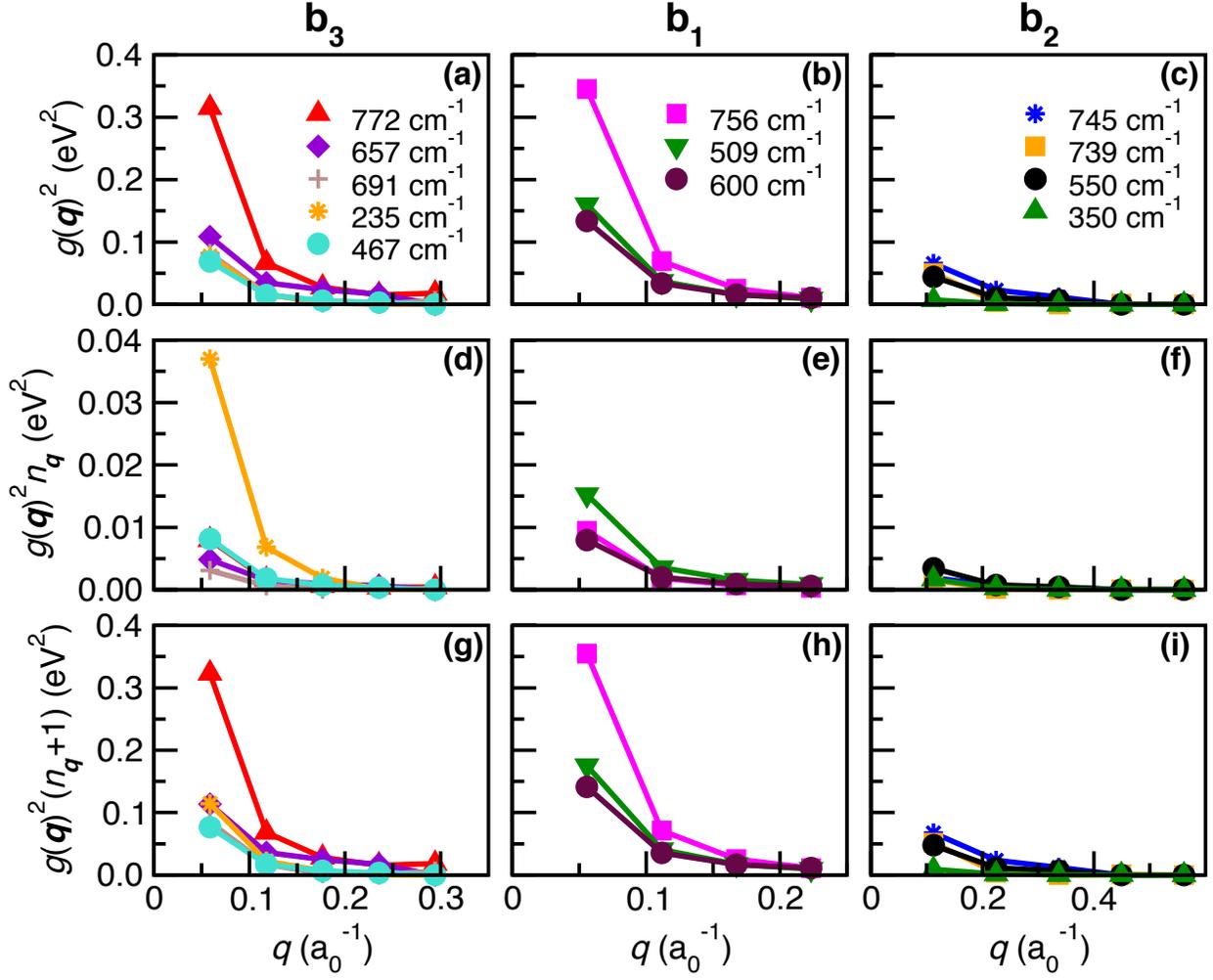

FIG. 5. The square of the intraband electron-phonon coupling matrix elements $g^2$ for the bottom conduction band of $\beta$-Ga$_2$O$_3$ between $\Gamma$ and $q$ along the three reciprocal lattice directions (a)-(c). The dominant modes for the phonon-absorption (d)-(f) and phonon emission (g)-(i) terms are also plotted for comparison. The legends denote the calculated phonon frequency for each mode at the wave vector included closest to $\Gamma$. Although the mode at 235 cm$^{-1}$ does not yield the strongest matrix element $g^2$, it dominates the phonon-absorption term ($g^2 n$) due to its low frequency (and associated large phonon occupation number) and limits the electron mobility at room temperature.

TABLE VI. The $C$ coefficients for the Fröhlich model for the dominant phonon modes are shown. The coefficients were calculated using the electron-phonon coupling matrix elements $g$ calculated at the wave vector closest to Γ for each direction ($0.1\times\boldsymbol{b}_j$) with Eq. (7).

| Reciprocal Lattice Direction | Phonon Frequency (cm$^{-1}$) | Fröhlich Model $C$ Coefficient (meV$^2$/a$_0^2$) |
|---|---|---|
| $\boldsymbol{b}_1$ | 261 | 40 |
|  | 275 | 65 |
|  | 509 | 498 |
|  | 600 | 416 |
|  | 756 | 1074 |
| $\boldsymbol{b}_2$ | 335 | 32 |
|  | 350 | 93 |
|  | 550 | 559 |
|  | 572 | 119 |
|  | 739 | 634 |
|  | 745 | 826 |
| $\boldsymbol{b}_3$ | 235 | 268 |
|  | 467 | 238 |
|  | 657 | 376 |
|  | 691 | 282 |
|  | 772 | 1091 |

**B. Origin of mobility limit**

Our results for the electron-phonon coupling matrix elements enable us to identify the limiting factors to the room-temperature mobility and interpret experimental results. The modes that couple strongest to electrons along each direction are the highest-frequency ones (772 cm$^{-1}$ along $\boldsymbol{b}_3$, 756 cm$^{-1}$ along $\boldsymbol{b}_1$, and 745 cm$^{-1}$ along $\boldsymbol{b}_2$). Those modes dominate phonon emission and, as discussed in Section III.C, play an important role in determining the dielectric-breakdown properties. For low-field transport, however, the thermal energy of electrons ($\sim k_B T$) is insufficient to cause electron scattering by the emission of those high-frequency phonons. Hence we turn our attention to those phonon modes that couple strongly to electrons yet have an energy that is sufficiently low (on the order of $k_B T$ or less) that allows them to be emitted by thermal electrons. The low-energy modes also have large phonon occupation numbers that facilitate electron scattering by phonon absorption. Of all the modes that dominate electron-phonon coupling, the mode with frequency 235 cm$^{-1}$ has a phonon energy (29 meV) close to $k_B T$ at room temperature (26 meV). The next lowest-energy mode that couples strongly to electrons has a frequency of 350 cm$^{-1}$ (43 meV). Despite having a lower $g^2(\boldsymbol{q})$ value than many of the other dominant modes, the 29 meV mode exhibits the largest contribution to the phonon-absorption term of electron-phonon coupling out of all modes for all three directions. The value of the phonon-emission term for the 29 meV mode is also comparable to the dominant high-frequency modes, yet its low energy allows those phonons to be emitted by thermal electrons and cause scattering. Therefore, our electron-phonon coupling results identified this particular phonon mode at 235 cm$^{-1}$ (29 meV) to likely be one of the fundamental intrinsic limitations of the electron mobility of $\beta$-Ga$_2$O$_3$ at room temperature. Our results also show that the highest mode, despite having large electron-phonon coupling matrix elements, does not dominate electron-phonon scattering due to the small activation of this mode at room temperature.

Our results agree with previously reported values for the frequency of the PO phonon mode that limits the room-temperature mobility (21 meV and 44 meV).[13,15] Hall-effect measurements by Ma *et al.* found that, for doping densities < 10$^{18}$ cm$^{-3}$, the electron mobility is intrinsically limited by phonons to < 200 cm$^2$ V$^{-1}$ s$^{-1}$ at 300 K.[13] The authors also estimated the dimensional Fröhlich coupling constant ($\alpha_F$) to be almost three times stronger than that of GaN. The atomistic origin behind the stronger electron-phonon scattering in Ga$_2$O$_3$ is the low symmetry of the crystal structure, which hosts low-energy polar optical modes. The lack of such low-frequency modes leads to a higher intrinsic electron mobility in GaN than $\beta$-Ga$_2$O$_3$.

**C. Dielectric breakdown**

We applied our calculated electron-phonon coupling matrix elements to estimate the intrinsic breakdown electric field of β-Ga$_2$O$_3$ following to the first-principles methodology developed by Sun *et al*.[44] The breakdown-field value $V_{br}$ is estimated according to:

$$V_{br} = \max\left[\frac{\sqrt{3m}}{e}\sqrt{\frac{1}{\tau(E)}B(E)}\right], E \in \{\text{CBM}, \text{CBM} + E_g\} \tag{8}$$

for $E_{CBM} \leq E \leq E_{CBM} + E_g$, where $E_g$ is the experimental band gap of β-Ga$_2$O$_3$ (4.5 eV), $m$ is the electron mass, $e$ is the electron charge, $\tau(E)$ is the electron relaxation time, and $B(E)$ is the net rate of energy loss. $\tau(E)$ is given by Equation (3) in Ref. 44 and is evaluated by summing over the probabilities of electronic scattering to all possible final states (Fermi's golden rule) and represents the time it takes for a single electron to independently relax from a certain energy to another. $B(E)$ is given by Equation (6) in Ref. 44 and is calculated similarly to $\tau(E)$ but accounts for energy exchanges between electrons and phonons either through phonon absorption or emission. Complete details for this method are described in the work by Sun *et al*.[44] $V_{br}$ is calculated in an energy range equal to the energy of the band gap, starting from the bottom of the CBM to $E_g$ higher in the conduction band(s). This limit is set by assuming that once an electron reaches energies higher than the band gap, impact ionization will occur at a rate of 100%. This assumption holds for a material sample with infinite size. Practical devices, however, have a finite channel length, and thus not all electrons that reach this energy impact ionize within the time that they cross the device. Therefore, the breakdown field in devices can be higher than our estimate for the bulk intrinsic material, and its analysis requires impact ionization coefficients of electrons such as those calculated by Ghosh and Singisetti.[45] To determine the breakdown field, we applied our previously calculated GW band structure[12] and the fitted Fröhlich parameters $C$ (see Table VI) for the electron-phonon coupling matrix elements to evaluate the scattering time and energy loss rate as a function of the electron energy above the conduction band minimum, similar to Ref. 46. We assumed screening by a carrier concentration of $10^{16}$ cm$^{-3}$ and a temperature of 300 K. We considered the contributions by the dominant phonon modes, which altogether account for approximately 80% of the total magnitude of the electron-phonon interaction.

The argument of the max function on the right-hand side of Eq. (8) is plotted as a function of energy in Fig. 6. The maximum values of the function for electron energies between the conduction-band minimum and an energy equal to the gap higher (4.5 eV) are 4.8, 5.5, and 5.9 MV/cm along the $\boldsymbol{b_1}$, $\boldsymbol{b_2}$, and $\boldsymbol{b_3}$ directions, respectively, which correspond to a directionally-averaged breakdown field of 5.4 MV/cm. This result was obtained considering only the dominant LO phonon modes, which account for ~80% of the total electron-phonon coupling magnitude along each direction. Multiplying the estimated breakdown field by a factor of 1.25 to extrapolate and account for the total electron-phonon interaction magnitude, we arrive at a breakdown-field value of 6.8 MV/cm. This estimated value is smaller than the earlier estimate of 8 MV/cm by Higashiwaki *et al.* based on the universal trend established by wide-band-gap semiconductors and assuming a gap value of 4.9 eV,[3] which at the time was the accepted gap value of $\beta$-Ga$_2$O$_3$. However, the value predicted using this trend is 6.3 MV/cm once the correct room-temperature gap value of 4.5 eV from Ref. 2 is considered, which agrees with our calculated result within ~0.5 MV/cm. Our results, therefore, agree with the estimated range of values for the breakdown field of 6-8 MV/cm for a band-gap range of 4.5-4.9 eV reported in recent work by Sasaki *et al.*[10]

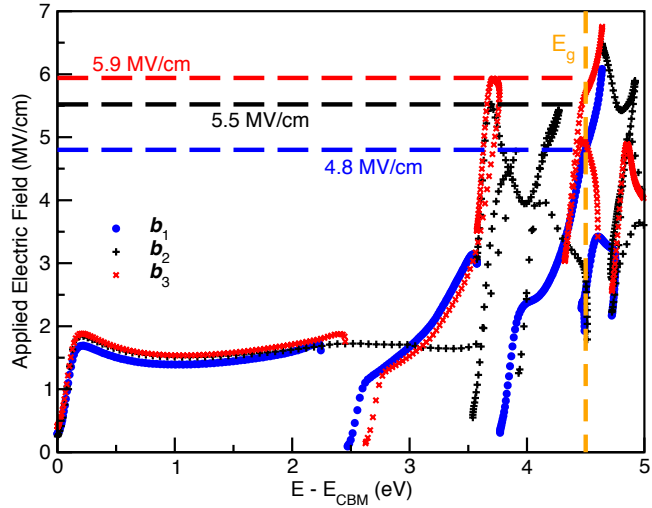

FIG. 6. Calculated estimate of the breakdown field of β-Ga$_2$O$_3$ along each reciprocal-lattice vector direction. The estimated values account only for electron scattering by the longitudinal optical phonon modes, which amount for ~80% of the total electron-phonon interaction. The experimental band-gap value of 4.5 eV is used for the upper limit of the electron energies. The calculated directionally averaged breakdown field is 5.4 MV/cm and is corrected to 6.8 MV/cm if extrapolated to account for the remaining 20% of the total electron-phonon interaction strength.

In a recent review, Stepanov *et al.*[6] used the empirical relationship developed by Hudgins *et al.*[47] to estimate the breakdown field of β-Ga$_2$O$_3$ and reported a value over 8 MV/cm. The authors used a band-gap value of 4.85 eV and the empirical relationship for a direct-gap material [$V_{br} = 1.73 \times 10^5 (E_g^{2.5})$] rather than an indirect-gap material [$V_{br} = 2.38 \times 10^5 (E_g^2)$], where $V_{br}$ has units of V/cm and $E_g$ is in eV, citing that the small energy difference between the direct and indirect gaps and the weakness of the indirect transitions essentially make $Ga_2O_3$ behave as a direct-gap material. If the band gap value of 4.5 eV is used, the Hudgins *et al.* empirical relationships predict breakdown fields of ~7.4 MV/cm (direct band gap) or ~4.8 MV/cm (indirect band gap), and the direct-gap estimate agrees well with our reported value of 6.8 MV/cm.

Our estimated breakdown field of $\beta$-$Ga_2O_3$ is lower than previous estimates and calls for a re-examination of the prospects of $\beta$-$Ga_2O_3$ for high-power electronic devices. Baliga's figure of merit (BFOM) depends on the breakdown field ($V_{br}$) to the third power at low frequencies (BFOM = $\epsilon \mu V_{br}^3$).[48,49] Previous estimates by Higashiwaki *et al.* assumed a breakdown field of 8 MV/cm and reported a BFOM at low frequencies for GaN and $\beta$-$Ga_2O_3$ as 870 and 3,444 relative to Si, respectively.[3] Our new estimate of 6.8 MV/cm reduces the BFOM for $\beta$-$Ga_2O_3$ to ~2,115 (~1.6 times lower) at low frequencies. Additionally, the value used for mobility in the previous report was 300 $cm^2$ $V^{-1}$ $s^{-1}$, which is actually an extrapolated experimental value.[5] However, recent reports in the literature place it in a lower range around ~200 $cm^2$ $V^{-1}$ $s^{-1}$.[13,50–53] Assuming a mobility of 200 $cm^2$ $V^{-1}$ $s^{-1}$ and a breakdown field of 6.8 MV/cm, the low-frequency BFOM of $Ga_2O_3$ is ~1,410 relative to Si, which is still 1.6 times greater than that of GaN. Despite the lower breakdown-field estimate (6.8 MV/cm) in combination with the lower mobilities recently measured, $\beta$-$Ga_2O_3$ still shows superior performance for power electronics compared to GaN.

Increasing the band gap even slightly by, e.g., alloying or strain, is a promising method to increase the breakdown field and the BFOM. From Fig. 6, for an energy equal to that of the conduction band minimum plus 4.5 eV (the experimental $E_g$), the average breakdown field is estimated to be 5.4 MV/cm, with values of 4.8, 5.5, and 5.9 MV/cm along the $\boldsymbol{b_1}$, $\boldsymbol{b_2}$, and $\boldsymbol{b_3}$ directions, respectively. Increasing the band gap to 4.7 eV causes a 20% increase in the average breakdown field estimate (6.5 MV/cm), with direction-dependent values of 6.1, 6.5, and 6.8 MV/cm along the $\boldsymbol{b_1}$, $\boldsymbol{b_2}$, and $\boldsymbol{b_3}$ directions. Accounting for the remaining 20% of the electron-phonon interaction would raise this value to ~8.1 MV/cm, bringing the breakdown field estimate to the original estimate of 8 MV/cm. Possible methods to realize the larger band gap include strain engineering or alloying with aluminum. The latter of which has already been demonstrated, and a band gap range from 5.2-7.1 eV was measured when using 24%-100% Al.[54] For applications in high-power electronics, future work should be done to improve the breakdown field of $\beta$-Ga$_2$O$_3$ by increasing the band gap as well as increasing the electron mobility and thermal conductivity.

## VI. CONCLUSIONS

In summary, we investigated the phonon properties, phonon-phonon interactions, and electron-phonon scattering of $\beta$-Ga$_2$O$_3$ in the anisotropic monoclinic crystal. We derived the directionally dependent phonon dispersion curves, LO-TO splittings, sound velocities and found good agreement with experiment. Oxygen substitution by O$^{15}$ has the largest isotopic effects on the phonon frequencies. Our calculated Grüneisen parameters indicate that the optical modes show stronger anharmonicities than the acoustic ones and suppress the thermal conductivity of $\beta$-Ga$_2$O$_3$. We also determined that the low-symmetry crystal structure gives rise to a low-frequency polar-optical mode with a phonon energy of 29 meV that dominates electron scattering at room temperature and limits the mobility. Our value for the breakdown field of $\beta$-Ga$_2$O$_3$ is 6.8 MV/cm, which is in good agreement with empirical estimates that use the revised band-gap value of 4.5 eV. We validate that $\beta$-Ga$_2$O$_3$ has a higher Baliga FOM compared to GaN. Our results identify the microscopic origins of the thermal and electron transport limits in $\beta$-Ga$_2$O$_3$ and propose strategies to increase the breakdown field to 8 MV/cm by increasing the band gap either by growth under strain or by alloying with Al$_2$O$_3$.

## ACKNOWLEDGEMENTS

This work was supported by the Designing Materials to Revolutionize and Engineer our Future (DMREF) Program under Award No. 1534221, funded by the National Science Foundation. K.A.M. acknowledges the support from the National Science Foundation Graduate Research Fellowship Program through Grant No. DGE 1256260. This research used resources of the National Energy Research Scientific Computing Center, a DOE Office of Science User Facility supported by the Office of Science of the U.S. Department of Energy.

Poncé, Y. Pouillon, T. Rangel, G.-M. Rignanese, A.H. Romero, B. Rousseau, O. Rubel, A.A. Shukri, M. Stankovski, M. Torrent, M.J. Van Setten, B. Van Troeye, M.J. Verstraete, D. Waroquier, J. Wiktor, B. Xu, A. Zhou, and J.W. Zwanziger, Comput. Phys. Commun. **205**, 106 (2016).

[28] T. Onuma, S. Saito, K. Sasaki, K. Goto, T. Masui, T. Yamaguchi, T. Honda, A. Kuramata, and M. Higashiwaki, Appl. Phys. Lett. **108**, 101904 (2016).

[29] D. Machon, P.F. McMillan, B. Xu, and J. Dong, Phys. Rev. B **73**, 094125 (2006).

[30] G. Kresse and J. Hafner, Phys. Rev. B **47**, 558 (1993).

[31] G. Kresse and J. Furthmüller, Phys. Rev. B **54**, 11169 (1996).

[32] G. Kresse and J. Furthmüller, Comput. Mater. Sci. **6**, 15 (1996).

[33] T. Onuma, S. Fujioka, T. Yamaguchi, Y. Itoh, M. Higashiwaki, K. Sasaki, T. Masui, and T. Honda, J. Cryst. Growth **401**, 330 (2014).

[34] J.B. Khurgin, D. Jena, and Y.J. Ding, Appl. Phys. Lett. **93**, 032110 (2008).

[35] K. Sarasamak, S. Limpijumnong, and W.R.L. Lambrecht, Phys. Rev. B **82**, 035201 (2010).

[36] X.L. Chen, J.K. Liang, Y.P. Xu, T. Xu, P.Z. Jiang, Y.D. Yu, and K.Q. Lu, Mod. Phys. Lett. B **13**, 285 (1999).

[37] F. D. Murnaghan, *Finite Deformation of an Elastic Solid* (Dover, New York, 1967).

[38] L. Vočadlo, J.P. Poirer, and G.D. Price, Am. Mineral. **85**, 390 (2000).

[39] V.Y. Vaschenko and V.N. Zubarev, Sov. Phys. Solid State **5**, 653 (1963).

[40] A.R. Goñi, H. Siegle, K. Syassen, C. Thomsen, and J.-M. Wagner, Phys. Rev. B **64**, 035205 (2001).

[41] M. Higashiwaki, K. Sasaki, H. Murakami, Y. Kumagai, A. Koukitu, A. Kuramata, T. Masui, and S. Yamakoshi, Semicond. Sci. Technol. **31**, 034001 (2016).
33